\documentclass[%
superscriptaddress,
reprint,
 amsmath,amssymb,
 aps,prl
]{revtex4-2}

\usepackage{graphicx}
\usepackage{dcolumn}
\usepackage{bm}
\usepackage{multirow}

\begin{document}

\title{Geometric and Topological Entropies of Sphere Packing }
\author{Jack A. Logan}
\affiliation{Department of Physics and Astronomy, Stony Brook University, Stony Brook, NY 11794}
\affiliation{Center for Functional Nanomaterials, 
Brookhaven National Laboratory, Upton NY 11973}
\author{Alexei V. Tkachenko}
\affiliation{Center for Functional Nanomaterials, 
Brookhaven National Laboratory, Upton NY 11973}
\begin{abstract}
We present a statistical mechanical description of randomly packed spherical particles, where the average coordination number is treated as a macroscopic thermodynamic variable. The overall packing entropy is shown to have two contributions: geometric, reflecting statistical weights of individual configurations, and topological, which corresponds to the number of topologically distinct states.  Both of them are computed in the thermodynamic limit for isostatic packings in  2D and 3D, and the result is further expanded to the case of ``floppy" particle clusters. The theory is directly applicable to sticky colloids, and in addition,  generalizes concepts of granular and glassy configurational entropies for the case of non-jammed systems.
\end{abstract}

\maketitle

The deceivingly simple task of packing spheres has been an inspiration for multiple  problems in mathematics and  natural sciences since the times of Kepler \cite{Hales2006}. It  provides  insights into the physics of crystalline solids, as well as  into the  world of disordered states of matter  \cite{bernal1959geometrical,torquato2000random,Parisi2005}. One such fundamental problem is finding the entropy of a random packing of spheres. In its simplified form, this  amounts to counting the number of ways in which distinct particle arrangements  can be generated for a given ensemble. For instance, more than two decades ago, Sam Edwards introduced the notion of granular  entropy \cite{edwards1989theory, edwards1990flow}.   A similar concept  also   arises in the context  of hard-sphere glasses, as a measure of degeneracy of the locally stable configurations \cite{Parisi2005,parisi2010mean}. In both examples, the packing entropy would be a measure of the multiplicity of the ``jammed" states. Due to the non-equilibrium nature of jamming, even a rigorous definition of the granular entropy remains  non-trivial.  Nevertheless,  significant progress in understanding and computing it  has  been demonstrated  in recent years  \cite{song2008phase,briscoe2008entropy,asenjo2014numerical,martiniani2016turning}. 

In this paper, we discuss the packing entropy in an equilibrium  system of spheres, where it can be properly defined as a conventional thermodynamic quantity. The system studied is not jammed, but rather  is constrained to have a specific number of direct  interparticle contacts. In other words, we use the total number of contacts (or equivalently,  mean coordination number of the particles, $Z$) as a thermodynamic variable. This approach is immediately relevant to packings  of ``sticky" spheres where each contact is associated with a fixed binding energy. Two limits of that problem have been explored in the past: thermodynamics of a sticky sphere liquid (e.g. Baxter model) \cite{baxter1968percus,miller2004phase}, and, more recently, free energy landscapes and kinetics of mesoscopic multi-colloidal clusters \cite{meng2010free,holmes2013geometrical,perry2015two,holmes2017sticky,cates2015celebrating,klein2018physical}. In this work, we seek to bridge the gap between these two limits, and,  more importantly, use this model system to better understand  the nature of packing entropy. The latter can be subsequently connected to the granular and/or glassy entropy by applying external pressure to the system that would lead to its jamming. The difference from the original contexts in which those entropies were introduced, is that one  would start with an already discrete configurational space, and select its  subset that corresponds to the local minima of the total volume.  

Consider a system of $N$ hard spherical particles in $d$-dimensional space, with average particle diameter $\bar a$. The packing is weakly polydisperse, so  that the width of the particle diameter distribution is much smaller than the average: $\delta a \ll \bar a$. We define a pair of particles to be in contact if the gap between them is less than some small value $\Delta\ll \bar a$. The gap between particles $i$ and $j$ is defined as  $x_{ij}=\left| {\bf r}_i-{\bf r}_j\right|-(a_i+a_j)/2$, where $a$'s and ${\bf r}$'s are their respective diameters and positions. For any configuration, the topology of the packing can be specified by an adjacency matrix $\hat{\bf C}$, with elements $C_{ij} = 1$ for all particles $i \neq j$ in contact, and $C_{ij} = 0$ otherwise. As  already mentioned above, the average coordination number $Z=1/N\sum_{i<j}C_{ij}$ is treated as a macroscopic thermodynamic variable of the system. Note that throughout the paper we set $k_B T = 1$. 

The weak polydispersity is introduced to avoid hyperstatic (over-constrained) configurations. These contain  ``accidental" contacts that   could be removed, e.g.,  by slight variations of particle sizes (subject to the constraint that all other contacts are intact). Examples of such over-constrained configurations are close packed crystals (FCC/HCP in 3D or hexagonal lattice in 2D). If those are disqualified, any rigid  packing has to be isostatic, i.e. the number of contacts has to be equal to the total number of degrees of freedom of $N$ spheres, $dN$, minus the number of rigid body degrees of freedom of the packing as a whole, $d(d+1)/2$ \cite{Alexander_PhysRep1998,AT1999}. In the thermodynamic limit, this corresponds to $Z^*=2d$, while for a finite isostatic  packing $Z^*=2d-d(d+1)/N$.  Below we start by discussing the isostatic limit, and then generalize our results  to  the under-constrained case $Z<Z^*$, in which each ``missing"  bond gives rise to a single zero mode.

The {\it packing entropy} for a given $Z$ is found by performing the summation of statistical weights of all topologically distinct  realizations:
\begin{equation}
\label{eq:pack}
        e^{NS_{pack}(Z)}= \sum_{\bf \hat C}\frac{\delta(Z({\bf \hat C})-Z)}{N!} e^{ N S_{geo}({\bf \hat C})}
\end{equation}
Here we have introduced the {\it geometric entropy} $S_{geo}({\bf \hat C})$ which determines the statistical weight of a specific  realization of the packing:
\begin{equation}
    \label{geom}
    e^{NS_{geo}({\bf \hat C})}=\int {\frac{\mathrm{d}^d{\bf r}_2...\mathrm{d}^d{\bf r}_{N}}{
    \Omega_d\bar{a}^{(N-1)d}}\prod_{i>j, C_{ij}=1}{\bar{a}\delta(x_{ij})}  \prod_{i>j, C_{ij}=0}{\Theta(x_{ij})}}
\end{equation}
This expression assumes that all $N$ spheres  belong to a single cluster.
Division by the factor  $\Omega_d$ eliminates contributions from its rigid body rotations  (specifically,  $\Omega_2= 2\pi$ for  2D and $\Omega_3=2\pi \cdot 4\pi=8\pi^2$ for 3D). Translation of the cluster as a whole is not included since integration is only performed over positions on $N-1$ out of $N$ spheres. Without loss of generality, the position of the first particle ${\bf r}_1$ will be assumed to be fixed at the origin. The factor $N!$ in Eq.~(\ref{eq:pack}) deserves a special clarification due to the widespread confusion regarding its origin in statistical mechanics. As we show in Supplementary Materials (SM), this factor does not require the particles to be indistinguishable, and moreover,  it has nothing to do with quantum mechanics \cite{asenjo2014numerical,cates2015celebrating}.

For an  isostatic packing, we can switch variables from the positions $({\bf r}_2,...,{\bf r}_N)$ to the gaps between pairs of particles in contact, $(x_1,...,{x}_{NZ/2})$. The new variables should also include  $d(d-1)/2$ independent rigid body  rotations of the cluster:  ${\bf \hat  \theta}=\left(\theta_1...,\theta_{d(d-1)/2}\right)$.  The Jacobian   associated with  this change in variables is  known as the rigidity matrix  \cite{holmes2013geometrical}. According to Eq.~(\ref{geom}), the corresponding Jacobian determinant can be used to find the geometric entropy:
\begin{equation}
   \label{eq:logjacobian}
  NS_{geo}({\bf \hat C})= \ln\left[ \frac{\partial\left({\bf r}_2,...,{\bf r}_N\right)}{\partial\left({\bf \hat  \theta},x_1,...,{x}_{NZ/2} \right)} \right]
\end{equation}

A more conventional approach to finding the statistical weight for each topologically distinct configuration, e.g. in the context of sticky colloids,  is to calculate  the free energy that includes both phonons and rigid body modes of the entire packing \cite{perry2015two,holmes2013geometrical,meng2010free,holmes2017sticky,cates2015celebrating}. That route is practical, but it has led to a number of seemingly paradoxical observations. In particular, one needs to assign specific masses to all of the particles, and replace rigid bonds with effective springs. Of course, in the non-quantum regime masses may only give a constant contribution to the free energy. And yet, the overall statistical weight of a cluster in this formulation depends, e.g. on its moment of inertia \cite{cates2015celebrating}. The paradox may be  resolved by the direct demonstration that when the phonon and rotational partition functions are combined, all of the masses would only give rise to a trivial multiplier, independent of specific configurations \cite{klein2018physical}. The same is true for the spring constants: as expected in the isostatic  case, the free energy is not  sensitive to the  details of the bonding potential \cite{holmes2017sticky}.  The geometric entropy gives the statistical weights of any  configuration in the form of Eq. (\ref{eq:logjacobian}), making the phonon-based calculation redundant.


\begin{figure}[ht]
\includegraphics[width=0.7\columnwidth]{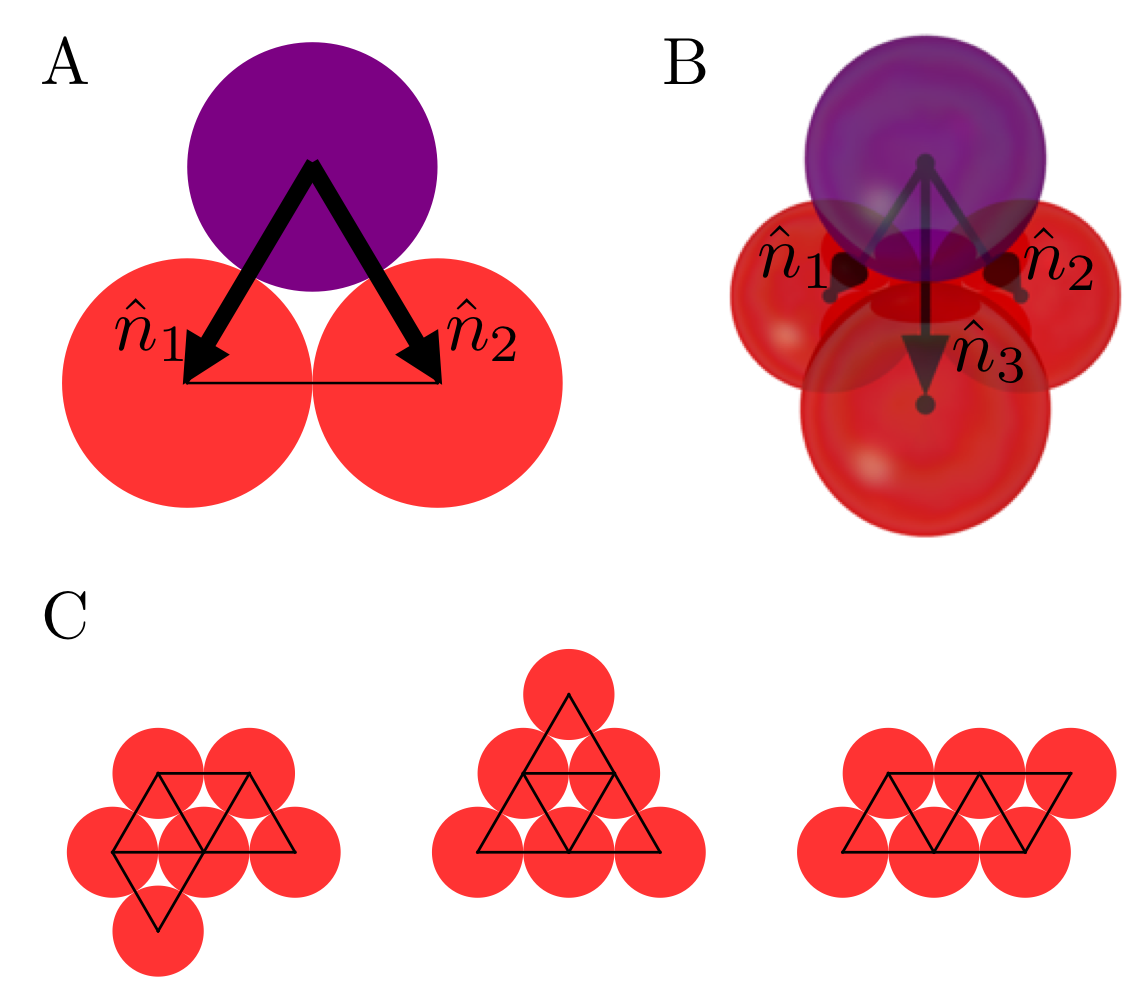}
\caption{(A,B) Minimal clusters (red) for $d=2$ and $d=3$. An additional particle (purple) bonds to each particle in the minimal cluster along vectors $\hat{n}_i$. (C) Three six-particle clusters with the same $S_{geo}$, despite having different topologies.}
\label{fig:minimal_clusters} 
\end{figure}

Consider some examples of calculating the geometric entropy. We define a minimal cluster in $d$ dimensions as $d$ particles, all in contact with each other,  such as the red particles in Fig.~\ref{fig:minimal_clusters}A-B. A minimal cluster is a special case that gives $S_{geo}=0$. The proof for $d=2,3$ is presented in SM. The next most trivial cluster is formed by adding a particle that bonds to each of the particles in the minimal cluster, such as the purple particles shown in the figure. In 2D, the geometric entropy associated with this  additional particle is  $S_{geo}=\ln \left[\hat{n}_1\times \hat{n}_2\right]$, where $\hat{n}_i$ are unit vectors in the directions of the bonds. If the particles are arranged in a square lattice, for example, the bond directions $\hat{n}_1$ and $\hat{n}_2$ are orthogonal, and hence  $S_{geo}=0$. In the case of a triangular lattice, however, each additional particle forms an equilateral triangle and adds $-\mathrm{ln}(\frac{\sqrt{3}}{2})$ to the overall geometric entropy of the minimal cluster. Similarly in 3D, adding a particle to the minimal cluster creates a regular tetrahedron, and the triple product of the unit vectors along the bonds adds $-\mathrm{ln}(\frac{1}{\sqrt{2}})$ to the geometric entropy. Just like the square lattice in 2D, a  cubic lattice is a special case of an isostatic packing with $S_{geo}=0$ because the bonds are mutually orthogonal. 

We can continue to add particles to the minimal clusters that have exactly $d$ bonds each, maintaining the isostaticity of the packing. Three such examples where four particles have been added to a 2D minimal cluster are shown in Fig.~\ref{fig:minimal_clusters}C. The addition of each new particle always increases the cumulative geometric entropy of the cluster by $-\mathrm{ln}(\frac{\sqrt{3}}{2})$. For this reason, despite the differences in their topologies, they all have the same $S_{geo}$. Note, however, that if the distinction between particles is ignored, the higher-symmetry triangular cluster would have a {\it lower} statistical weight, as it corresponds to a smaller number of non-trivial particle permutations \cite{perry2015two,klein2018physical}.



The overall packing entropy can be expressed as 
 \begin{eqnarray}
  \label{eq:spack_geotopo}
  S_{pack}(Z) =\langle S_{geo}\rangle_{Z}+S_{topo}(Z) 
\end{eqnarray}
Here $\langle ... \rangle$ denotes the averaging over all topological realizations, weighted proportionally to $\exp(NS_{geo})$, with a given $Z$. The fact that $-N S_{geo}$ acts  as an effective Hamiltonian  enables  one to employ a Monte-Carlo (MC) approach to generate the equilibrium ensemble of isostatic packs and calculate, e.g. $\langle S_{geo}\rangle_{Z}$ itself, or any other ensemble-averaged quantity. However, the generated ``energy" landscape is very rough, with lots of local minima, and making large, non-local, moves to escape them  is highly  non-trivial.  To resolve this complication, we introduce a generalized effective  Hamiltonian,  $H_{eff} = - \lambda N S_{geo}(\mathbf{\hat{C}})$. The parameter $\lambda$ here allows one to tune the model from its original form (for $\lambda=1$), to one with a completely flat ``energy" landscape (in $\lambda=0$ limit). That limit corresponds to a model in which  all plausible topological arrangements with the same coordination number $Z$ have the same statistical weight.  MC simulations are run for  $\lambda=0$, and  the results for $\lambda=1$ are extrapolated from  those  simulations  by calculating  the  linear response to  $\lambda$. Let $S_{geo}^{(0)}={\langle S_{geo}\rangle}$  and  $S^{(0)}_{pack}$ be the geometric and the full packing  entropies computed for $\lambda=0$ (note that topological  entropy in the limit  $\lambda=0$ is identical to $S^{(0)}_{pack}$). By expanding $S_{pack}$ in leading orders  of $\lambda$ and $S_{geo}$,  we obtain:   

\begin{equation}
\label{eq:S_vs_Sgeo}
    S_{pack}(\lambda, S_{geo}) = \lambda S_{geo} + S^{(0)}_{pack} - \frac{1}{2\chi}\left(S_{geo} - S^{(0)}_{geo}\right)^2
\end{equation}
Here, the linear response coefficient  $\chi=N\left\langle(S_{geo} -  S^{(0)}_{geo})^2 \right\rangle$ can be extracted from statistics of $S_{geo}$.  By maximizing Eq. (\ref{eq:S_vs_Sgeo}) with respect to $S_{geo}$, one can  extrapolate  geometric,  topological, and the overall packing  entropy to  $\lambda=1$. 
In particular, that gives $S_{geo}\approx  S^{(0)}_{geo} +\chi$ and  $S_{topo}^* = S^{(0)}_{pack} - \frac{\chi}{2}$. Below we describe the numerical procedure that was used  to compute $S_{geo}^{(0)}$, $\chi$ and  $S^{(0)}_{pack}$.

\begin{figure}[h!]
    \centering
     \includegraphics[width=0.8\columnwidth]{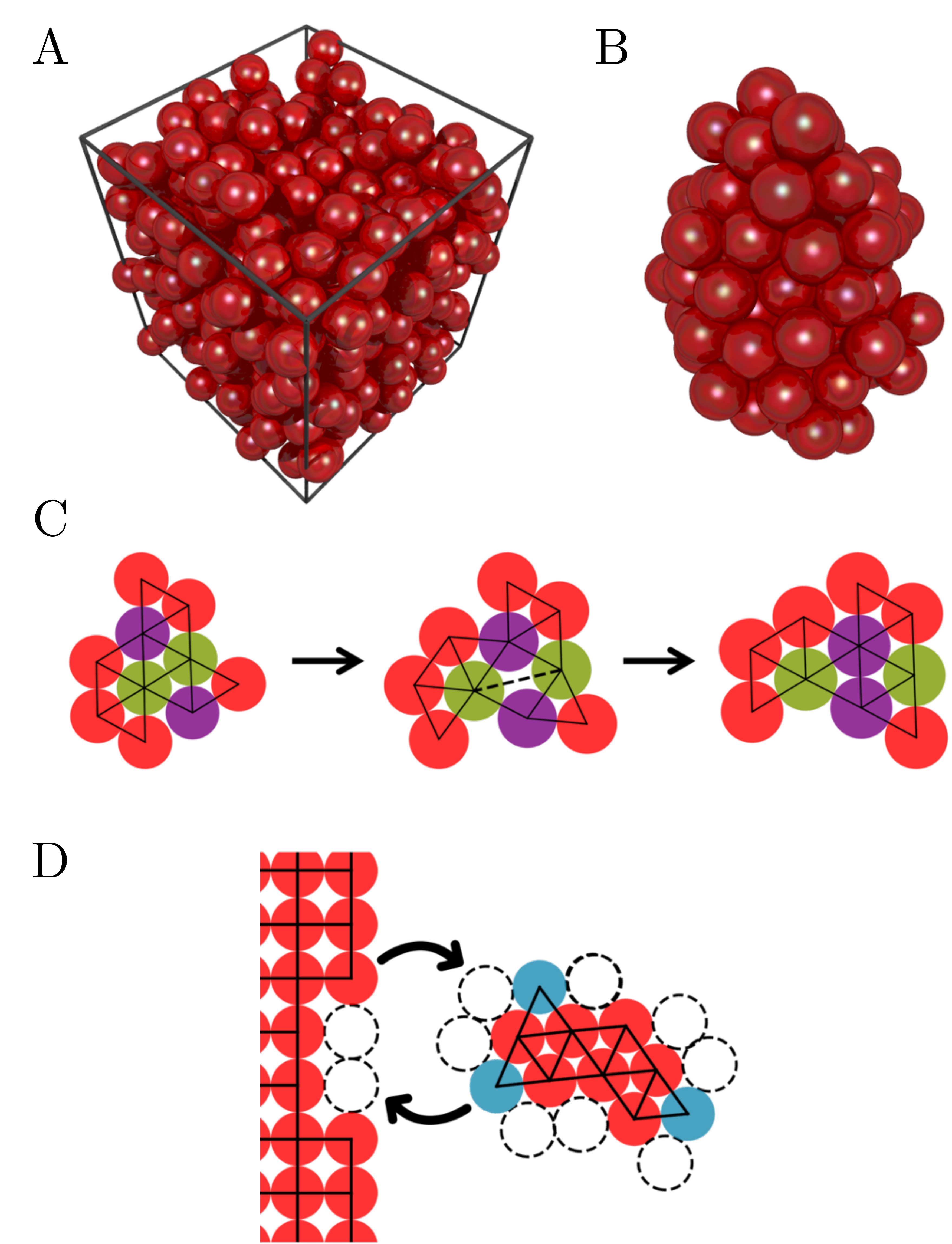}
    \caption{Example packings. (A) Packing with periodic boundary conditions in the horizontal directions, with a closed bottom, but open upward. (B) A typical cluster packing in free space. (C) A typical Monte Carlo move. The bonded green particles slowly open, allowing the packing to move along a single zero mode until the purple particles make contact. At this point, the bond is broken between the green pair of particles and created between the purple particles, completing the move to a new topological state. (D) A cluster packing in equilibrium with an infinitely large square lattice. At any moment, a free particle may ``condense" onto the packing in any of the $N_+$ positions where it will have exactly $d$ bonds (dashed open circles). Likewise, any of the $N_-$ particles in the packing with exactly $d$ bonds (blue) may ``evaporate" off of the packing and back to the lattice. Here $N_+ = 9$ (the top-center is two closely spaced) and $N_- = 3$.}
    \label{fig:packings_and_MC_moves}
\end{figure}

In our simulations, it was ensured  that the packing remains  strictly  isostatic and does not have any interparticle overlaps.  A typical MC move is illustrated in Fig.~\ref{fig:packings_and_MC_moves}C: a randomly chosen bond is broken, and the gap between the two particles  increases, while all the  spheres are pushed along the single zero mode associated with the lost contact. The move is complete once  two, previously unbound, particles make a contact. If $\lambda = 0$, there is no difference in statistical weights between different (isostatic) configurations, so the only constraint is non-overlapping of the particles.

Two different classes of isostatic systems have been studied: (i) packings in a semi-periodic box, with periodic boundary conditions (PBC) along the x- and y- axes, and a free boundary in the upward z-direction (only x- and z- axes in 2D), and (ii) clusters in open space. In both cases, the size of the system was gradually increased, by adding one particle at a time, in a way that preserved the overall isostaticity.     Examples of both packing types can bee seen in Fig.~\ref{fig:packings_and_MC_moves}A-B. 








By using Eq.(\ref{eq:logjacobian}), distribution  of $S_{geo}$ has been  computed directly, which determined parameters $S_{geo}^{(0)}$ and $\chi$ in Eq. (\ref{eq:S_vs_Sgeo}).        In order to compute  $S_{pack}^{(0)}$, we imagine that the packing can exchange particles with an infinitely large cubic (or square in 2D) lattice through ``evaporation" and ``condensation" events, as shown in Fig.~\ref{fig:packings_and_MC_moves}D. Technically, particles have not been moved between the random  packing and the reference lattice, but only the probabilities of such moves have been computed. In equilibrium,    chemical potential  of  particles in the random packing, $\mu=-S_{pack}^{(0)}$, should match that in the reference lattice.  For any configuration, we find a number $N_+$ of sites where a particle can be added to the packing, and a number $N_-$ of ``removable" particles that has exactly $d$ bonds. By requiring the addition and removal processes  to balance each other,  one obtains   $S_{pack}^{(0)}=-\mu=\ln \left(\langle N_+ \rangle / \langle N_- \rangle  \right)$.

 \begin{figure}[ht]
    \centering
      \includegraphics[width=0.9\columnwidth]{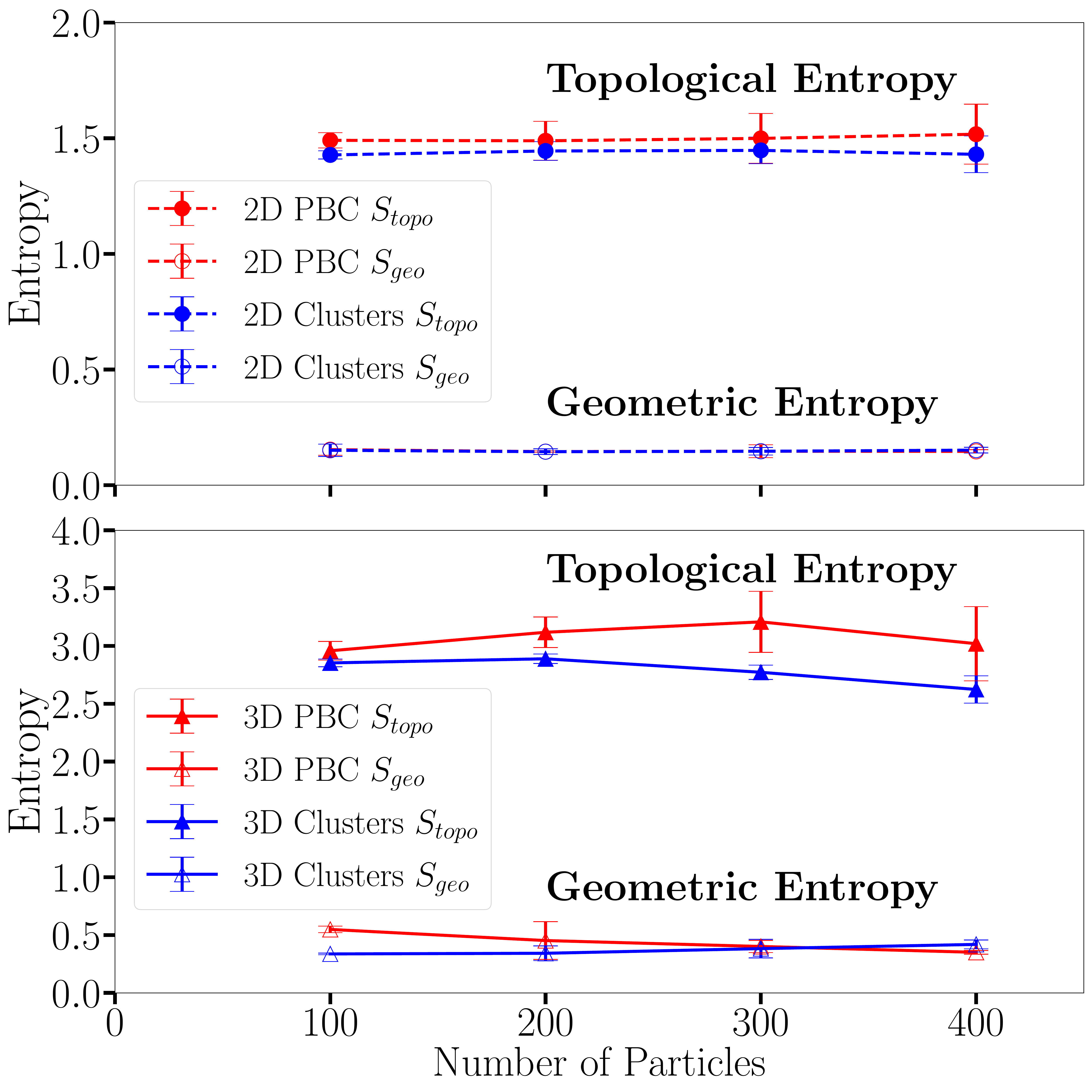}
    \caption{Geometric and Topological entropy values per particle for 2D (top, dashed) and 3D (bottom, solid), including PBC packings (red) and cluster packings (blue) for all packing sizes, up to 400 particles.}
    \label{fig:Stopo_and_Sgeo_vs_N}
\end{figure}

Our results for the topological and geometric entropies are shown  in Fig.~\ref{fig:Stopo_and_Sgeo_vs_N} as functions of packing size for both cases of PBC and clusters. Note an excellent agreement between the two methods, and  apparent convergence in the limit of large $N$.  The implied  infinite-size  values of different types of entropies for 2D and 3D, together with the corresponding volume fractions $\eta$,   are presented  in Table~\ref{table:entropy_values}.  Interestingly, the obtained geometric entropies are quite close to our original  estimates  which were based on a sequential packing procedure: $S_{geo}\approx 0.15\pm 0.01$, vs. $-\mathrm{ln}(\frac{\sqrt{3}}{2})\approx 0.14$ in 2D, and   $S_{geo}\approx 0.35\pm 0.04$ vs. $-\mathrm{ln}(\frac{1}{\sqrt{2}})\approx 0.35$ in 3D. Topological entropies in both cases are quite substantial: $1.52 \pm 0.13$ and $3.02\pm 0.32$, respectively. It is important to keep in mind that we are exploring the space of isostatic packs, rather than jammed ones. In other words, they do not represent local density maxima. This explains why our results for the  topological entropy are significantly larger than those extracted from recent computation of granular entropies: $0.5$ and $0.7$ for 2D and 3D, respectively \cite{asenjo2014numerical,martiniani2016turning}. This also explains why the observed equilibrium volume fraction in 3D is significantly lower than that of  Random Close Packing (or,  Maximally Random Jammed state defined in Ref.  \cite{torquato2000random}). Curiously, the found equilibrium value $\eta\approx 0.54\pm 0.02$  is close to the so-called random loose packing density \cite{loose}, although this is likely to  be  just a coincidence.  

\begin{table}[ht]
\centering
\begin{tabular}{|c|c|c||c|} 
\hline
\multicolumn{2}{|c|}{} & \textbf{PBC} & \textbf{Clusters} \\ 
\hline \hline 
\multirow{4}{*}{$\mathbf{d=2}$} & $\mathbf{S_{pack}}$ & 1.66 & 1.58  \\ \cline{2-4}
& $\mathbf{S_{geo}}$ & 0.15 & 0.15 \\ \cline{2-4}
& $\mathbf{S_{topo}}$ & 1.52 & 1.43 \\
\cline{2-4}
& $\boldsymbol{\eta}$ & 0.73 & 0.78 \\ 
\hline \hline
\multirow{4}{*}{$\mathbf{d=3}$} & $\mathbf{S_{pack}}$ & 3.37 & 3.04  \\ \cline{2-4}
& $\mathbf{S_{geo}}$ & 0.35 & 0.42 \\ \cline{2-4}
& $\mathbf{S_{topo}}$ & 3.02 & 2.62 \\ \cline{2-4}
& $\boldsymbol{\eta}$ & 0.52 & 0.56 \\ 
\hline
\end{tabular}
\caption{The geometric, topological, and full packing entropy per particle in 2D and 3D Cluster and PBC packings with 400 particles. The average packing fraction $\langle \eta \rangle$ is also listed.}
\label{table:entropy_values}
\end{table}

So far, we focused on an isostatic case, where the number of contacts  exactly matched the number of frozen degrees of freedom in a rigid packing. To expand our approach to the under-constrained  case of so-called  ``floppy" networks, we note that breaking any bond in an isostatic network generates a single zero mode. Such a  mode is bounded by two isostatic states, which allows one to count the number of distinct underconstained configurations, and find the corresponding topological entropy:
\begin{equation}
 S_{topo}(Z)\approx 
    S_{topo}(Z^*)+\frac{Z^*-Z}{2}\ln\left[\frac{ Z^*}{2(Z^*-Z)}\right]
\end{equation}
 
 As for the geometric entropy, it is obtained by integration over all activated zero modes, and the corresponding correction is simply  proportional to their number: $S_{geo}(Z)\approx S_{geo}(Z^*)+\frac{Z^*-Z}{2}\ln(\xi/\bar{a})$, where $\xi$ represents  the typical range over which the interparticle gap $x_{ij}$ may change until the new contact is formed. In the specific case of sticky spheres, the packing entropy $S_{topo}+S_{geo}$ makes an important contribution to the chemical potential of a disordered aggregate (i.e. a large  cluster):
\begin{equation}
\mu(Z)=\frac{\epsilon Z}{2} - S_{geo}(Z)-S_{topo}(Z)
\end{equation} 
Here  $\epsilon=-\ln\left(\int_0^{\infty}\exp(-V(x))dx/\bar{a}\right)$ is the binding free energy, determined by short-range interparticle potential $V(x)$.  Minimization of chemical potential $\mu(Z)$ predicts an  exponential suppression of zero modes  with bond strength: $(Z^*-Z)\sim e^{-\epsilon}$. Therefore, residual topological and geometric entropies of an aggregate are close to their isostatic values, given in  Table~\ref{table:entropy_values}. 

In summary, we proposed a  statistical mechanical description of sphere packings  based on treating the  coordination number $Z$ as a macroscopic thermodynamic parameter, and  identified two contributions to the packing entropy: geometric and topological. They correspond to the statistical weight of a particular topological configuration, and the number of non-equivalent arrangements, respectively. The topological entropy is thus analogous, but not equivalent, to Edwards granular entropy or to the residual entropy of hard sphere glasses. An important difference of our approach is that it is built entirely within the framework of equilibrium statistical mechanics, and does not impose a requirement on the individual configurations to be jammed.  Hence, our results are directly applicable to systems with tensile short-range forces, such as sticky spherical colloids.    We developed an MC scheme to compute the geometric and topological entropies of isostatic packings in both 2D and 3D, and further generalized the results for the case of floppy (under-constrained)  packings.




{\bf Acknowledgment.} The research was carried out at  the Center for Functional Nanomaterials, which is a U.S. DOE Office of Science Facility, at Brookhaven National Laboratory under Contract No. DE-SC0012704

\bibliography{packing_letter}
\newpage
\newpage
\setcounter{page}{1} 
 \newcounter{sfigure}
  \renewcommand{\thefigure}{S\arabic{sfigure}}
  \newcounter{stable}
  \renewcommand{\thetable}{S\arabic{stable}}
  \newcounter{sequation}
  \renewcommand{\theequation}{S\arabic{sequation}}
  \stepcounter{sfigure}
  \stepcounter{stable}
  \stepcounter{sequation}
  
\section{Supplementary Materials}
\subsection{Origin of factor $N!$}
\label{app:origin_of_N_factorial}
There is considerable confusion in statistical physics literature (including multiple textbooks) regarding  the  origin of the $N!$ factor that appears in Eq. (\ref{eq:pack}). Traditionally, this factor is introduced  as a resolution to the Gibbs paradox, with a justification of the particles (e.g.,  molecules of a gas) being ``indistinguishable". It is often claimed that the origin of this factor lies in quantum mechanics. Neither of these justifications is valid. Statistical mechanics of an  ideal gas is routinely applied e.g. to systems of colloids which are   neither quantum nor strictly identical. To clarify this issue, consider a system, containing $N$ particles that can be exchanged with a much larger system (thermal bath, TB) containing $N_{tot}-N$ particles. All of the particles in this example are distinguishable, but we assume the free energy of TB to only depend on the total number of particles that it contains, not on their specific subset (e.g. the thermal bath may be an ideal gas). If we start with all particles being in the thermal bath, there are $N_{tot}$ ways of selecting a particle to be moved  to the system,  then there are $N_{tot}-1$ ways of selecting the next one, etc. When taking into account the arbitrary order in which $N$ particles can be moved from TB to the system, we obtain the overall statistical weight of the configuration with a given  $N$:
\begin{align}
   \nonumber\tilde{Z}(N)=\lim_{N_{tot}\rightarrow \infty}\frac{N_{tot}!}{N!(N_{tot}-N)!} \frac{Z(N)Z_{tb}(N_{tot}-N)}{Z_{tb}(N_{tot})} =\\ =\frac{Z(N)e^{-\mu N}}{N!}
\end{align}
 Here  $\mu =-\partial  \ln (Z_{tb}(N_{tot}) )/\partial N_{tot} - \ln N_{tot}$ is  chemical potential of a particle in the thermal bath. Partition function of the system $Z(N)$ is averaged over all possible selections of $N$ particles from $N_{tot}$ that belong to the TB.

\subsection{Geometric Entropy of Minimal Clusters}
\label{app:minimal_clusters_zero_sgeo}
        
            \begin{figure}[ht]
                \centering
              \includegraphics[width=0.5\columnwidth]{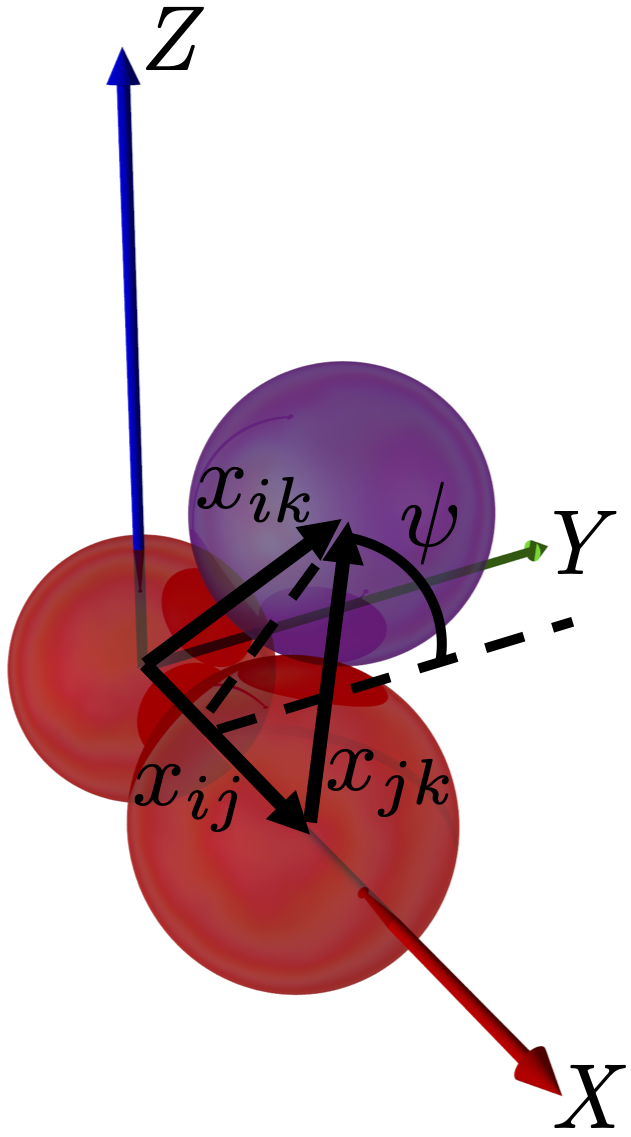}
                \caption{Initial two particles ($i$ and $j$) shown in red along the $X$-axis in contact with particle $k$ (purple). Particle $k$ has a rotational degree of freedom, here seen at an angle $\psi$ with the $XY$-plane, that allows it to remain in contact with the other particles.}
                \label{fig:sgeo_0_min_clusters}
        \end{figure}
        \stepcounter{sfigure}

        Consider two identical particles, $i$ and $j$, of diameter $a$. Particle $i$ is fixed as the origin of the system, and particle $j$ is placed along the $X$-axis in contact with $i$, as shown in Fig.~\ref{fig:sgeo_0_min_clusters}. To get the Jacobian,
        we consider infinitesimal displacements $\delta \mathbf{r}$,  $\delta \boldsymbol{\theta}$, $\delta \boldsymbol{\phi}$. The effect of $\delta \mathbf{r}$ is to change the gap $x_{ij}$ between the particles, while the other two displacements lead to small changes in $y_j$ and $z_j$ through infinitesimal rotations. Altogether we find
        \begin{gather}
            \begin{bmatrix}
                \delta x_j \\
                \delta y_j \\
                \delta z_j
            \end{bmatrix} = 
            \begin{bmatrix}
                1 & 0 & 0 \\ 
                0 & a & 0 \\ 
                0 & 0 & -a
            \end{bmatrix}
            \begin{bmatrix}
                \delta r \\
                \delta \theta \\
                \delta \phi
            \end{bmatrix}.
        \end{gather}
        \stepcounter{sequation}
        
        The Jacobian determinant, given $a=1$, is then $|J|=1$. This serves two purposes: it allows us to build to a 3D minimal cluster one particle at a time, and restricting above to the $XY$-plane shows that a 2D minimal cluster has geometric entropy of zero.
        
        For a minimal cluster in 3D, we add a third identical particle $k$ to the configuration above. The two previous particles are fixed along the $X$-axis, and the third particle is placed in contact with them. We need to transform coordinates from $(x_k, y_k, z_k)$ to $(x_{ik}, x_{jk}, \psi)$, where $\psi$ is the angle from the $XY$-plane to particle $k$, and $x_{ik}$ and $x_{jk}$ are the gaps between the particles. Again we consider infinitesimal displacements, this time in $\delta \mathbf{x_k}$, $\delta \mathbf{y_k}$, and $\delta \mathbf{z_k}$ and write down how it affects the gap sizes $x_{ik}$, $x_{jk}$, and the tilt angle $\psi$.
        
             \begin{gather}
            \begin{bmatrix}
                \delta x_{ik} \\
                \delta x_{jk} \\
                \delta \psi
            \end{bmatrix} = 
            \begin{bmatrix}
                \frac{1}{2} & 0 & \frac{\sqrt{3}}{2} \\ 
                -\frac{1}{2} & 0 & \frac{\sqrt{3}}{2} \\ 
                0 & -\frac{2}{\sqrt{3}} & 0
            \end{bmatrix}
            \begin{bmatrix}
                \delta x_k \\
                \delta y_k \\
                \delta z_k
            \end{bmatrix}
        \end{gather}
        \stepcounter{sequation}
        
        We find the Jacobian determinant is again $|J|=1$.
        For 3D, the product of the jacobians for each individual particle gives the complete Jacobian value. The entropy is then found as the natural logarithm of the Jacobian, $S_{geo} = \ln(1) = 0$.

\subsection{Packing Evolution}

A packing, whether PBC or cluster, evolves in a $dN$-dimensional space that can be represented by the positions of the $N$ spheres or the $dN$ bond gaps. The constraints for them to be hard spheres in contact are given by $|\vec{r}_i - \vec{r}_j|^2 = \left(\frac{a_i + a_j}{2}\right)^2$. The Jacobian of this equation is the rigidity tensor, shown in Eq.~(\ref{eq:logjacobian}). At each step of our simulations a bond is broken and a zero mode enters into the packing. The packing then evolves in such a way that it moves orthogonal to the constraints imposed by the rigidity tensor until contact is made between two unbonded particles, and a new isostatic packing is realized. The rigidity tensor $R(\vec{r})$ relates the particle displacements $\vec{u}_i$ to the gaps between particles $x_{ij}$. If we consider that the bond $\alpha$ breaks and opens by an amount $x_{\alpha}$, the displacement of any particle $i$ can be calculated as $\vec{u}_{i} = R^{-1}_{i,\alpha} x_{\alpha}$. However, to avoid the computationally expensive inverse function, we instead use a QR decomposition with column pivoting \cite{eigenweb} to solve the equation for the displacements $\vec{u}_i$ of all particles caused by the opening of $\alpha$. 

\begin{figure}[h]
    \centering
    \includegraphics[width=1\columnwidth]{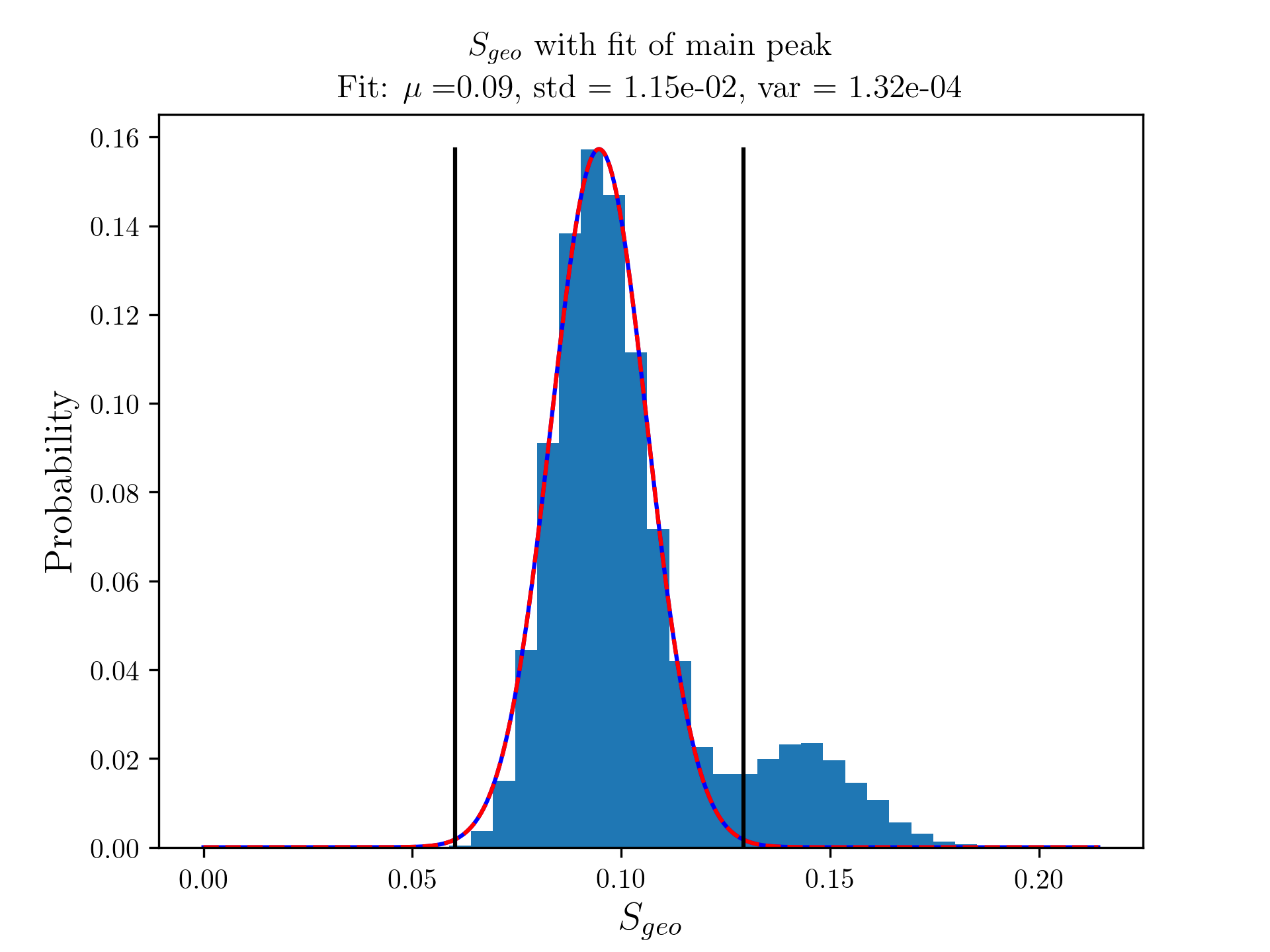}
    \caption{The original $S_{geo}$ distribution for the $400$-particle PBC packings. A clear second peak is seen that points to zero modes persisting in the configurations after a new topology is made. Fit with a normal distrubtion in dashed red. The black vertical lines represent $3\sigma$ of the fit.}
    \label{fig:sgeo_w_zero_modes}    
\end{figure}
\stepcounter{sfigure}

It's important that the trajectory of the evolution moves the packing from one isostatic packing to another. The new bond that should be made is completely determined by the specific bond that is broken. Sometimes, however, due to the finite precision of the simulation, a bond other than the ``true" bond will close. When this occurs one part of the packing will be overconstrained, while a zero mode will persist in another. This zero mode appears as anomalously high geometric entropy in our distribution, as demonstrated by the second peak that appears at larger values than the main peak in Fig.~\ref{fig:sgeo_w_zero_modes}. This peak is a collection of packings that were made when the incorrect bond was closed. In order to maintain results that only include true isostatic packings, we must filter out the configurations that include zero modes. We first fit our main peak with a normal distribution, shown in red in the figure. The black vertical lines mark three standard deviations from the mean of the fit. Any data outside of the black lines is assumed to come from a zero mode configuration and is discarded.

\subsection{Simulation  Procedure}
\label{app:simulation_details}
We begin a PBC simulation with a fixed layer of particles at the base of the box ($z=0$). This layer acts as a substrate on which particles can adsorb. These substrate particles are fixed throughout the simulation, but any bonds they share with free particles are not. For the packing to be isostatic, it requires $dN_{free}$ bonds to match the number of translational degrees of freedom from the free particles. We use Eq.~(\ref{eq:logjacobian}) to calculate the geometric entropy for each configuration.

To build a packing, we first identify every possible location where a new particle could come into existence and have exactly $d$ bonds without leading to any overlaps with the current particles. We call these positions ``virtual particles". In general, for any configuration there exists an enormous number of virtual particles, and we randomly choose one in a way that tends to maximize our packing fraction by applying a weight to each virtual particle. While building the packing to the desired number of particles, our MC move (Fig.~\ref{fig:packings_and_MC_moves}C) is used to rearrange the packing, allowing it to mix while it grows. The results are not particularly meaningful while the number of particles is constantly changing, so no data is collected while a packing is growing.

\begin{figure}[ht]
    \centering
    \includegraphics[width=1\columnwidth]{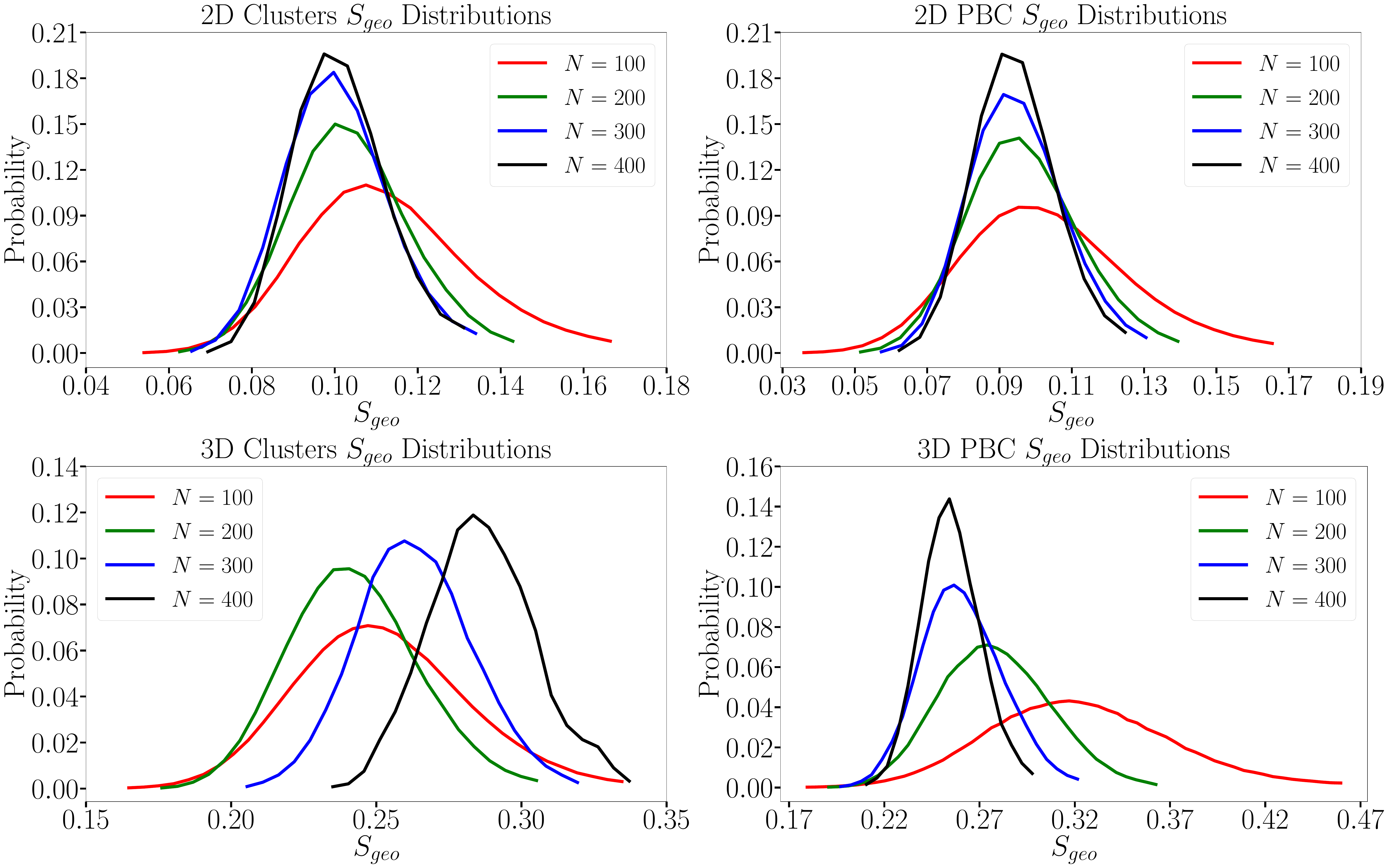}
    \caption{The distributions of geometric entropy for 2D clusters and PBC packings (top) and 3D clusters and PBC packings (bottom) computed using our MC algorithm. These distributions represent the geometric entropy for different configurations with $\lambda=0$.}
    \label{fig:Sgeo_distributions}    
\end{figure}
\stepcounter{sfigure} 

When all of the boundaries are free, as they are for a cluster, both the translational and the rotational degrees of freedom need to be considered. For $d=3$, for instance, Maxwell's criterion for stiffness tells us that there must be exactly $3N-6$ bonds. In order to do this, for any given configuration of $N$ total particles, we fix $d$ bonded particles that are randomly chosen. Simply fixing $d$ particles leaves the system underconstrained, so $d$ bonds between the particles must also be fixed. For this reason, it is imperative that we choose $d$ particles that are all bonded to each other---a minimal cluster, as defined in the main text. For $d=3$ the minimal cluster can be any rigid triangle of three particles with three bonds, and for $d=2$ it can be any bonded pair of particles (Fig.~\ref{fig:minimal_clusters}). In this way, we find ourselves in a similar situation to a PBC packing---the number of contacts considered exactly matches the number of degrees of freedom. This allows us, again, to use Eq.~(\ref{eq:logjacobian}) to find the geometric entropy. In this case, however, it is broken into two parts: the Jacobian for the $N-d$ particles that are free to move, analogous to the PBC packing, and an additional Jacobian for the $d$ particles in the minimal cluster that have been chosen to be temporarily fixed. The Jacobian of the minimal cluster is calculated to be proportional to the product of the distances between the $d$ fixed particles in contact. The full Jacobian to be used in Eq.~(\ref{eq:logjacobian}) is the product of the Jacobian of the $N-d$ free particles and that of the $d$ fixed particles.


\begin{figure}[ht]
    \centering
    \includegraphics[width=1\columnwidth]{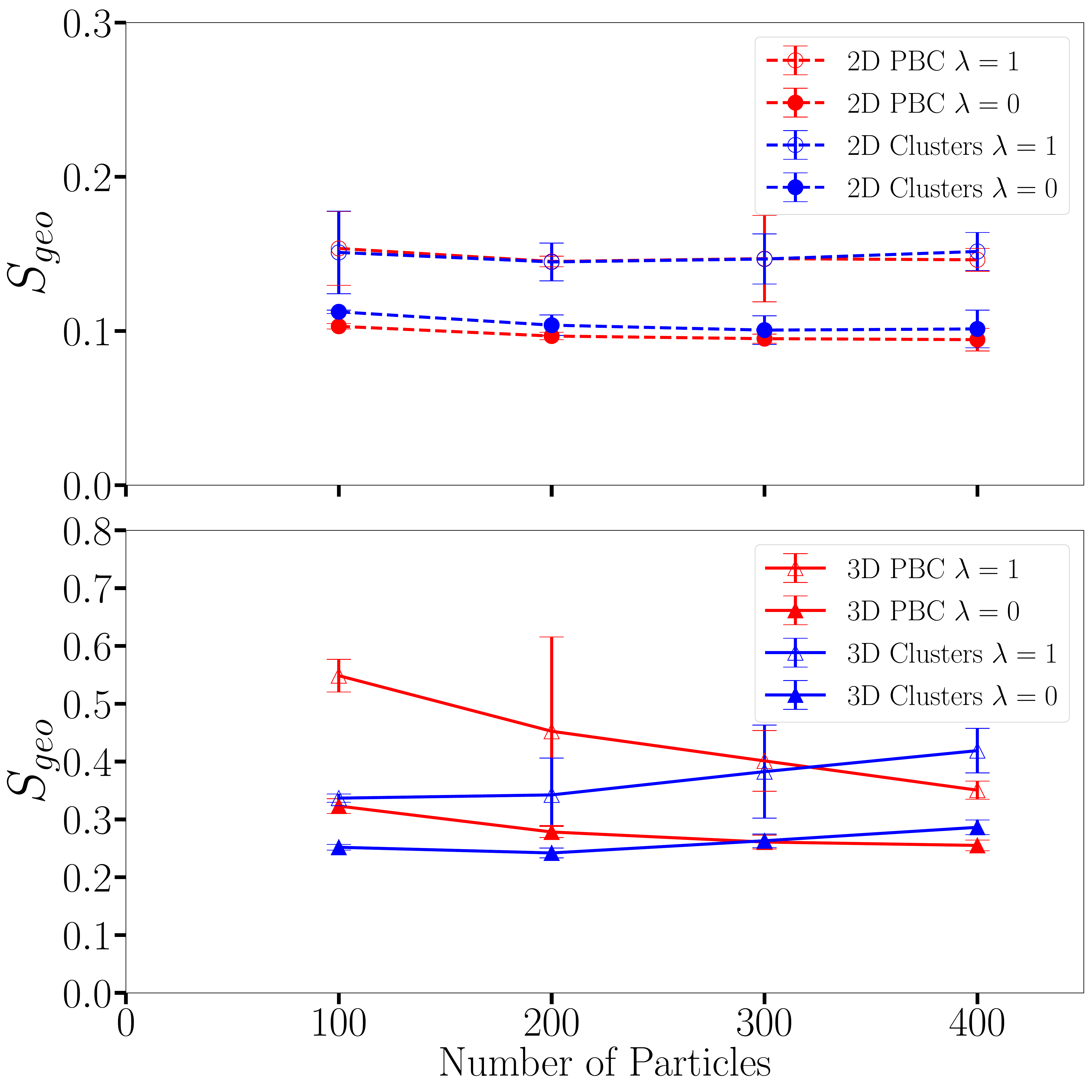}
    \caption{The geometric entropy for all packing sizes in 2D (top) and 3D (bottom). The values for $\lambda=0$ are drawn with solid markers, while the extrapolated $\lambda=1$ values have open markers. PBC packings are shown in red and cluster packings in blue.}
    \label{fig:Sgeo_vs_N}
\end{figure}
\stepcounter{sfigure}

    
     




\end{document}